\documentclass[twocolumn,showpacs,preprintnumbers,amsmath,amssymb]{revtex4-1}

\usepackage{graphicx}
\usepackage{dcolumn}
\usepackage{bm}

\begin{document}

\preprint{APS/XX0000}

\title{Beta Decays of Isotones with Neutron Magic Number of $N$=126 and
R-process Nucleosynthesis} 

\author{Toshio Suzuki$^{1,2,3}$, 
Takashi Yoshida$^{4}$,
Toshitaka Kajino$^{3,4}$,
and Takaharu Otsuka$^{5,6}$}
\affiliation{$^{1}$ Department of Physics, College of Humanities and
Sciences, Nihon University\\     
Sakurajosui 3-25-40, Setagaya-ku, Tokyo 156-8550, Japan\\
$^{2}$ Center for Nuclear Study, University of Tokyo, Hirosawa, Wako-shi, 
Saitama 351-0198, Japan\\
%
$^{3}$ National Astronomical Observatory of Japan, 
Mitaka, Tokyo 181-8588, Japan\\
$^{4}$ Deaprtment of Astronomy, Graduate School of Science, 
University of Tokyo, Bunkyo-ku, Tokyo 113-0033, Japan\\
$^{5}$ Department of Physics and Center for Nuclear Study, University of Tokyo, Hongo, 
Bunkyo-ku, Tokyo 113-0033, Japan\\
$^{6}$ National Superconducting Cyclotron Laboratory, Michigan State University,
East Lansing, Michigan, 48824, USA}


\date{\today}

\begin{abstract}

Beta decays of the isotones with $N$=126 are studied by shell model 
calculations taking into account both the Gamow-Teller (GT) and 
first-forbidden (FF) transitions.
The FF transitions are found to be important to reduce the half-lives, by
nearly twice to several times, from those by the GT contributions only.
Possible implications of the short half-lives of the waiting point nuclei 
on the r-process nucleosynthesis during the supernova explosions are 
discussed. A slight shift of the third peak of the element abundances in 
the r-process toward higher mass region is found.

\end{abstract}

\maketitle


\def\be{\begin{equation}}
\def\ee{\end{equation}}
\def\bea{\begin{eqnarray}}
\def\eea{\end{eqnarray}}
\def\br{\bf r}


    



\section{INTRODUCTION}

The r-process is the most promising process for the synthesis of about 
a half of heavy elements beyond iron\cite{RM}.
Study of the r-process element synthesis has been done by considering
neutrino-driven winds in supernova explosions \cite{MMH}
as well as ONeMg supernovae \cite{Wan03} and
neutron-star meregers \cite{Thi}.
The production of the r-process elements in collapsars relating to long 
gamma-ray bursts has also
been investigated \cite{Fuji06}.
Recent observations of r-process elements in extremely metal-poor stars 
\cite{Honda} have suggested that the r-process has occured in explosions of massive
stars or in neutron-star mergers from early stage of Galactic chemical 
evolution.
The r-process is affected by various inputs of nuclear properties such as
mass formulae \cite{Wan04}, $\beta$-decay rates, (n, $\gamma$) 
and ($\gamma$, n) reaction cross sections, reaction rates of $\alpha$-processes
in light nuclei \cite{Kaj}, etc., and depends also on 
astrophysical conditions; electron-to-baryon number ratio, $Y_{e}$, 
the entropy and temperature of the explosion environment, 
and neutrino processes \cite{Mey,GLM}.
The evaluation of $\beta$-decay rates, particularly at the waiting point nuclei,
is one of the important issues of the nucleosynthesis through the r-process.
Investigations on the $\beta$-decays of isotones with neutron magic number of 
$N$=82 have been done by various methods including shell model \cite{LP}, 
QRPA/FRDM \cite{Moll}, QRPA/ETFSI \cite{Bor00} 
and HFB+QRPA \cite{Eng} calculations as well as CQRPA ones \cite{Bor03}.
The half-lives of nuclei obtained by these calculations are rather consistent to
one another, and especially in shell model calculations experimental half-lives at
proton number of $Z$ =47, 48 and 49 are well reproduced \cite{LP}.   

For the $\beta$-decays at $N$=126 isotones, however, half-lives obtained by
various calculations differ to one another \cite{GLM}. 
First-forbidden (FF) transitions become important for these nuclei in 
addition to the Gamow-Teller (GT) transitions in contrast to the case of 
$N$=82.
A strong suppression of the half-lives for the $N$=126 isotones due to the
FF transitions has been predicted in Ref. \cite{Bor03}.
Shell model calculations of the $\beta$-decays of $N$=126 isotones have been 
done only with the contributions from the GT transitions \cite{Lang,GLM}.
Moreover, experimental data for the $\beta$-decays in this region of nuclei are
lacking. The region near the waiting point nuclei at $N$=126 is, therefore, called
'blank spot' region.

Here, we study $\beta$-decays of $N$=126 isotones by taking into account both
the GT and FF transitions to evaluate their half-lives.
Shell model calculations are done with the use of shell model interactions,
modified G-matrix elements, that reproduce well the observed energy levels of
the isotones with a few (2 to 5) proton holes outside $^{208}$Pb \cite{Stee,Ryd}.  
In sect. II, GT and FF transition strengths for the $N$=126 isotones are
evaluated, and half-lives of the $\beta$-decays are obtained by taking
into account both the transitions. 
The calculated half-lives of the waiting point nuclei in the r-process
become shorter by including the contributions from the FF 
transitions.
In Sect. III, possible implications of these half-lives for the r-process
nucleosynthesis are discussed. A summary is given in Sect. IV.   

\section{Transition Strengths and Half-lives of the $N$ =126 Isotones}
\subsection{Gamow-Teller, Spin-dipole Transition Strengths and
Shape Factors}

First, the method for the evaluation of the rates and half-lives of the GT
and FF transitions are explained. Then isotone dependence of the GT and FF 
transition strengths as well as the effects of the inclusion of the FF 
transitions on the half-lives of the isotones are discussed.  

The decay rates, $\Lambda$, as well as the partial half-lives, $t_{1/2}$, of the
transitions are obtained by the following formulae \cite{WB,Schop,TH};

\begin{eqnarray}
\Lambda  ={\rm ln}2/t_{1/2} =f/8896 \quad({\rm s}^{-1}) \nonumber\\
f = \int_{1}^{w_0} C(w) F(Z,w) pw (w_0-w)^2 dw \nonumber\\
C(w) = K_0 +K_1 w +K_{-1}/w +K_2 w^2, 
\end{eqnarray}
where $w$ is the electron energy, $F(Z, w)$ is the Fermi function, 
and $K_n$'s depend on nuclear transition matrix elements.   
Here, relativistic corrections from the expansion of electron radial wave
functions in powers of electron mass and nuclear charge parameters are 
included;  
matrix elements of one-body operators, 
[$\vec{r}\times\vec{\sigma}$]$^{\lambda}$ with $\lambda$ =0, 1, 2 and 
$\vec{r}$, as well as those from weak hadronic currents, $\gamma_5$, 
$\vec{\alpha}$, are taken into account for the FF transitions. 
In case of the GT transition, the shape factor $C(w)$ does not
depend on electron energy; 

\begin{eqnarray}
K_0 &=& \frac{1}{2J_i+1} |<f|| O(1^{+}) ||i>|^2 \nonumber\\
O(1^{+}) &=& g_A \vec{\sigma} t_{-}
\end{eqnarray}
with $g_{A}$ the axial vector coupling constant, $J_i$ is the spin 
of the initial state and $t_{-}|n>$ =$|p>$, and $K_n$ ($n\neq$0) =0.  

In case of the FF transitions, for 0$^{-}$ transitions,

\begin{eqnarray}
K_0 &=& \zeta_0^2 + \frac{1}{9}(M_0^S)^2 \nonumber\\
K_{-1} &=& -\frac{2}{3}\mu_1\gamma_1\zeta_0 M_0^S \nonumber\\
\zeta_0 &=& V + \frac{1}{3} M_{0}^{S} w_{0} \nonumber\\
V &=& M_{0}^{T} +\xi M_{0}^{S\prime} \nonumber\\
M_0^S &=& -g_A\sqrt{3}<f|| ir[C_1 \times \vec{\sigma}]^{0} t_{-}||i>C \nonumber\\
M_0^T &=&  -g_A <f||\gamma_5 t_{-}||i>C, 
\end{eqnarray}
where $\xi$ = $\alpha Z/2R$ with $\alpha$ the fine structure constant and 
$R$ is the  nuclear charge radius, $\gamma_1$ = $\sqrt{1-(\alpha Z)^2}$,
$C_{L}$ = $\sqrt{4\pi/(2L+1)}Y_{L}$, $C$ = 1/$\sqrt{2J_i +1}$.
The prime in $M_0^{S\prime}$ indicates that the effects of the finite nuclear charge 
distribution on the electron wave function are taken into account \cite{WB,BR}.      
For 1$^{-}$ transitions,
\begin{eqnarray}
K_0 &=& \zeta_1^2 +\frac{1}{9}(x+u)^2 -\frac{4}{9}\mu_1 \gamma_1 u(x+u) \nonumber\\
&+&\frac{1}{18}w_0^2 (2x+u)^2 -\frac{1}{18}\lambda_2 (2x-u)^2 \nonumber\\
K_1 &=& -\frac{4}{3}uY -\frac{1}{9}w_0 (4x^2+5u^2) \nonumber\\
K_{-1} &=& \frac{2}{3}\mu_1 \gamma_1 \zeta_1 (x+u) \nonumber\\
K_2 &=& \frac{1}{18} [8u^2 +(2x+u)^2 +\lambda_2 (2x-u)^2] \nonumber\\
\zeta_1 &=& Y +\frac{1}{3} (u-x) w_0 \nonumber\\
Y &=& \xi^{\prime}y -\xi (u^{\prime} +x^{\prime}) \nonumber\\
x &=& = -<f|| ir C_1 t_{-} ||i>C \nonumber\\
u &=& -g_A\sqrt{2} <f||ir [C_1 \times \vec{\sigma}]^{1} t_{-}||i>C \nonumber\\
\xi^{\prime}y &=& - <f|| \frac{i}{M_N}\vec{\nabla} t_{-} ||i>C, 
\end{eqnarray}
where $M_N$ is the nucleon mass and the primes in $x^{\prime}$ and $u^{\prime}$ 
indicate that the effects of finite nuclear charge distribution are taken 
into account \cite{WB,BR}.  
The quantities $\mu_1$ and $\lambda_2$ are defined \cite{Schop} in terms of 
electron wave functions and depend on electron momentum, $p_e$. 
Their values are close to unity; $\mu_1$ =0.9$\sim$1.0 and $\lambda_2$ =
0.7$\sim$1.0 at $p_e >$ 0.5 and gets larger than 1 at $p_e <$ 0.5 
for $Z$ considered here \cite{BJ}.   

For 2$^{-}$ transitions,
\begin{eqnarray}
K_0 &=& \frac{1}{12} z^2 (w_0^2 -\lambda_2) \nonumber\\
K_1 &=& -\frac{1}{6} z^2 w_0 \nonumber\\
K_2 &=&  \frac{1}{12} z^2 (1 +\lambda_2)\nonumber\\
z &=& 2g_A <f|| ir [C_1 \times \vec{\sigma}]^{2} t_{-}||i>C.
\end{eqnarray}

In the leading order, the FF transition operators are expressed as \cite{BM}

\begin{eqnarray}
C(w) &=& \frac{1}{2J_i+1} |<f|| O(\lambda^{-}) ||i>|^2 \nonumber\\
O(0^{-}) &=& g_A [\frac{\vec{\sigma}\cdot\vec{p}}{m} + \xi 
i\vec{\sigma}\cdot\vec{r}] t_{-} \nonumber\\
O(1^{-}) &=& [g_V\frac{\vec{p}}{m} - \xi
(g_A \vec{\sigma}\times\vec{r} -ig_V\vec{r})] t_{-} \nonumber\\
O(2^{-})_{\mu} &=& i\frac{g_A}{\sqrt{3}} [\vec{\sigma}\times\vec{r}]^2_{\mu}
\sqrt{\vec{p}_e^2 +\vec{q}_{\nu}^2} t_{-}
\end{eqnarray} 
where $g_V$ is the vector coupling constant.
Matrix elements of the first and the second parts of $O(0^{-})$, 
the first, second and the third terms of $O(1^{-})$, and $O(2^{-})$ correspond
to $M_{0}^{T}$ and $M_{0}^{S\prime}$, $\xi'y$, $u^{\prime}$ and $x^{\prime}$, 
and $z$ in Ref. \cite{WB}, respectively. 
The transition strengths constructed from the matrix elements of $O(0^{-})$ 
and $O(1^{-})$ in Eq. (6) are dominant parts of $K_{0}$, while that of 
$O(2^{-})$ in Eq. (6) becomes equivalent to Eq. (5) when $\lambda_2$ 
equals unity.

A shell model study was made for FF $\beta$ decays in the lead region,
$A$ =205$\sim$212, where an enhancement of the rank-zero matrix element of
$\gamma_{5}$, $M_0^{T}$, has been found \cite{W1}. The quenching of the rank-one and
rank-two components of the decay rate due to the core polarization effects
has been also investigated \cite{W1,W2}.
Here, the transition matrix elements and rates are evaluated by using 
Eqs. (1)$\sim$(5) following Refs. \cite{WB,Schop}.

The isotones with proton number of $Z$ =64$\sim$73, that is, nuclei
with $n_h$ =18$\sim$9 proton holes are considered here for the shell model
calculations\cite{OXB}.
A closed $N$ =126 shell configuration is assumed for the parent nucleus. 
Proton holes 
in $0h_{11}$, $1d_{3/2}$ and $2s_{1/2}$ orbits are taken into account
for the shell model calculations. 
Configurations with 2$\sim$4 and up to 2 holes are taken for $1d_{3/2}$ and 
$2s_{1/2}$ orbit, respectively. For the $0h_{11/2}$ orbit, ($n_h-$6) $\sim$
($n_h-$4) ($\sim$12 in case of $n_h$=17) hole configurations 
are taken into account for $n_h\leq$17. 
In case of $n_h$ =18 ($Z$ =64), 10$\sim$12 hole configurations for the $0h_{11/2}$
orbit and additional 0$\sim$2 hole configurations for the $1d_{5/2}$ orbit
are considered. 
For neutrons, the $0h_{9/2}$, $1f_{5/2,7/2}$, $2p_{1/2,3/2}$ and $0i_{13/2}$ 
orbits outside
the $N$ =82 core are taken as the model space, and in the $\beta$-decays a
neutron in these orbits changes into a proton whose orbit has holes.
The FF transitions are induced by $\nu 0i_{13/2}$ $\rightarrow$ $\pi 0h_{11/2}$ 
and $\nu (fp)$ $\rightarrow$ $\pi (sd)$ transitions while the GT transition
is dominantly induced by $\nu 0h_{9/2}$ $\rightarrow$ $\pi 0h_{11/2}$ transition.
In the $\beta$-decays, important contributions come from transitions to
low-lying states. The giant resonance region is energetically off the
$\beta$-decay windows even for the very neutron-rich cases. 
Though the calculations have been carried out with a restricted configurations,
important states
with low excitation energies are considered to be well described by the present
shell model calculations.

\begin{figure*}[tbh]
\hspace{-8mm}
\includegraphics[scale=0.73]{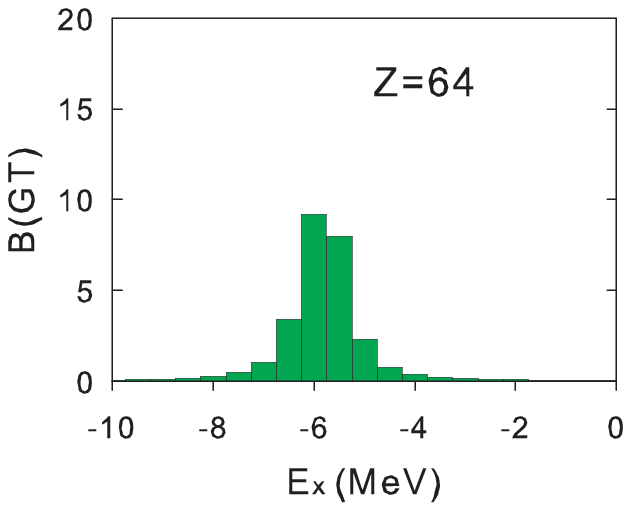}
\hspace{-11mm}
\includegraphics[scale=0.73]{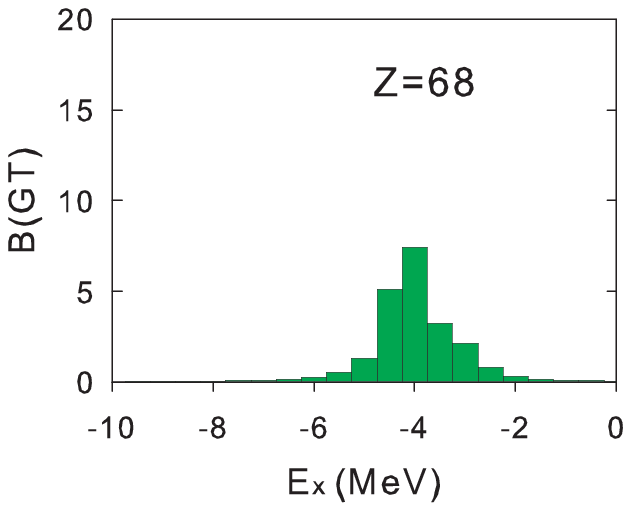}
\hspace{-10mm}
\includegraphics[scale=0.73]{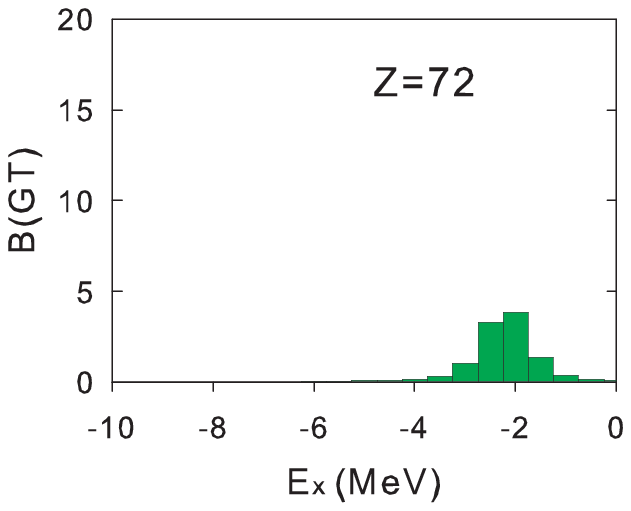}
\caption{(Color online) GT strengths for the isotone with $Z$=64, 68 
and 72  obtained by the shell model calculation at energies from the 
parent state.  
\label{fig:fig1}}
\end{figure*}

\begin{figure*}[tbh]
\hspace{-8mm}
\includegraphics[scale=0.72]{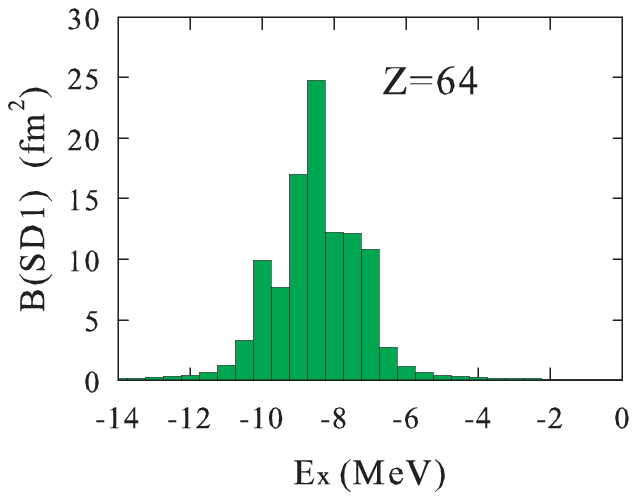}
\hspace{-11mm}
\includegraphics[scale=0.72]{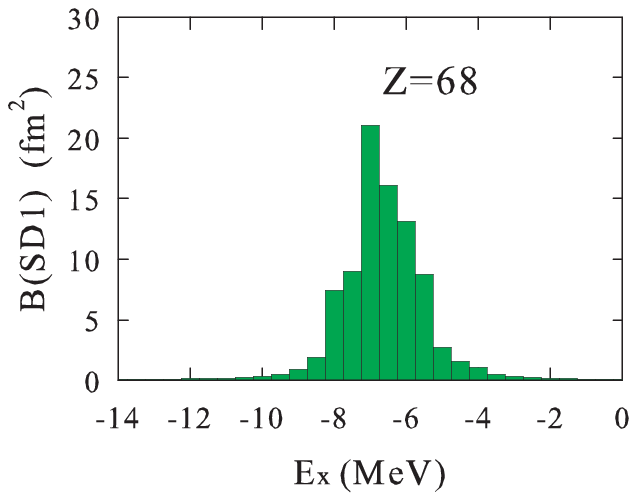}
\hspace{-10mm}
\includegraphics[scale=0.72]{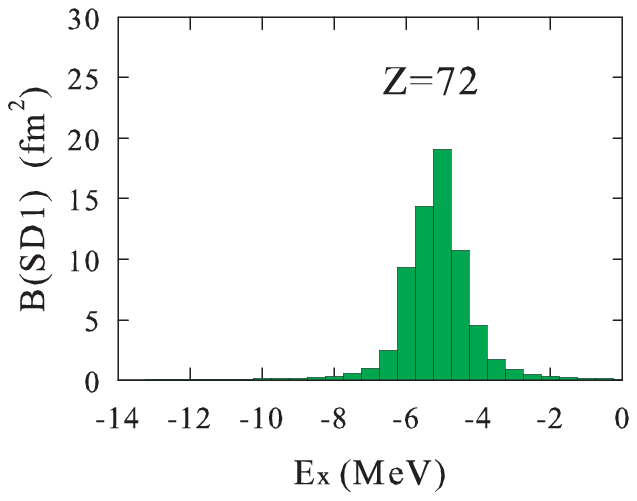}
%
\hspace{-20mm}
\includegraphics[scale=0.68]{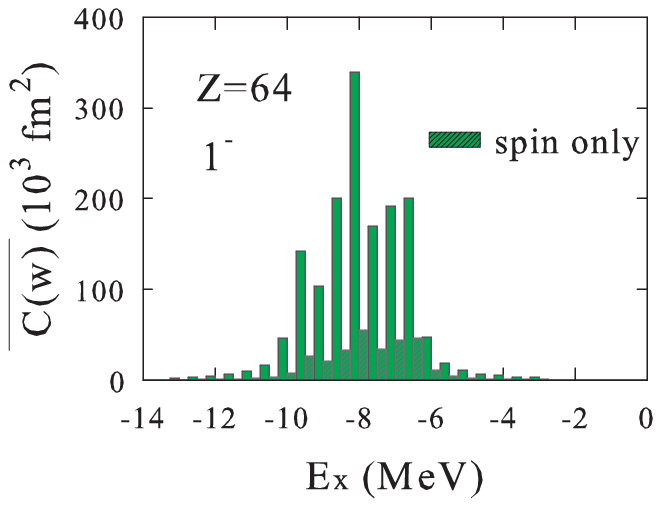}
\hspace{-11mm}
\includegraphics[scale=0.68]{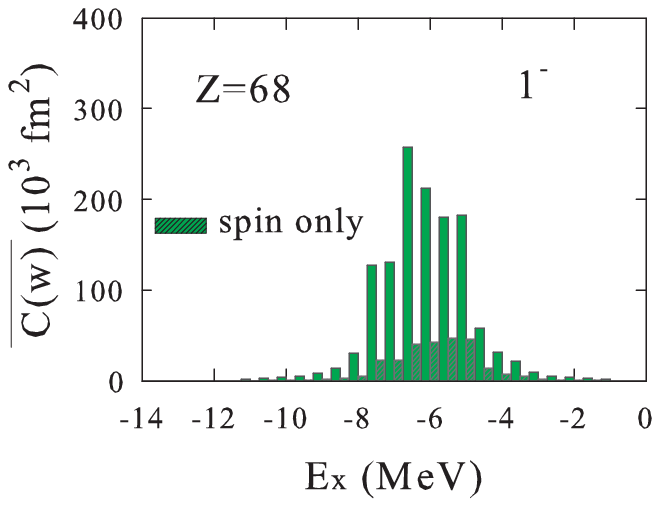}
\hspace{-11mm}
\includegraphics[scale=0.68]{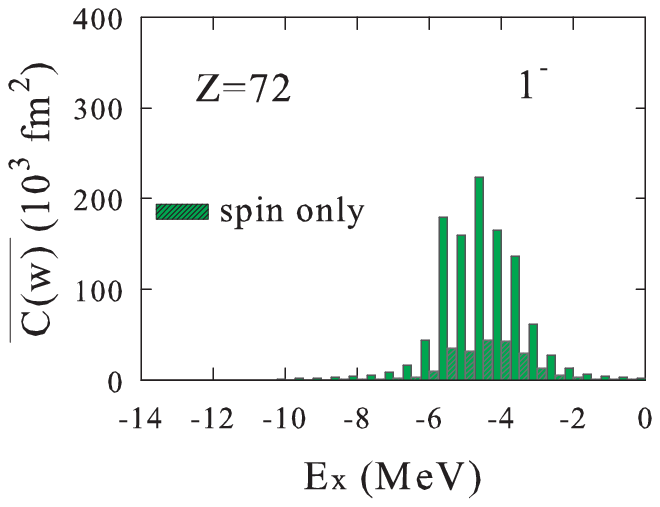}

\caption{(Color online) The same as in Fig. 1 for the spin-dipole strengths 
with $J^{\pi}$ =$1^{-}$ (upper figures) and the corresponding average shape 
factors (lower figures) for the isotones.
\label{fig:fig2}}
\end{figure*}

The GT transition strengths, $B(GT)$ = $K_0$ (eq. (2)), obtained by the shell 
model calculations are shown in Fig. 1. 
Here, $g_{A}$ is taken to be quenched by a factor of 0.7 \cite{Gar,Caur}, 
that is, $g_A$ =0.7$\times g_{A}^{free}$ with $g_{A}^{free}$=-1.26.   
Energies denoted are those from the parent state.
As the proton hole number increases, the energy difference between the parent
state and the daughter states at the peak of the strength and the phase 
space for the transition becomes larger. 
Note that the transition strength is approximately proportional to the fifth 
power of the energy difference. 
The summed GT strength also becomes generally larger for more proton holes; 
the sum of the calculated $B(GT)$ values are 14.4, 14.6, 11.7, 8.5 and 5.6
for $Z$ = 64, 66, 68, 70 and 72, respectively. Both of the effects lead to 
shorter half-lives for the isotones with larger proton hole number in the 
$\beta$-decays.       

We find that the most important contributions from the FF transitions come from 
the case of $J^{\pi}=1^{-}$. 
We discuss spin-dipole strengths which are important spin-dependent contributions
in $K_{0}$ as well as the total shape factors averaged over electron energy.
Calculated spin-dipoloe strengths 

\begin{equation}
 B(SD\lambda) = \frac{1}{2J_i+1}|<f||r [Y^{(1)}\times\vec{\sigma}]^{\lambda} t_{-} ||i>|^2
\end{equation}

\noindent with  $\lambda$=1 are shown in Fig. 2 (upper figures). 
General feature of the strength distributions is similar to the case of the GT
transitions. Energy difference between the parent state and the daughter states at 
the peak of the strength as well as the sum of the strength get larger for nuclei
with more proton holes.   
The summed $B(SD1)$ values are 55.5, 49.2, 48.4, 40.1 and 35.1 fm$^2$
for $Z$ =64, 66, 68, 70 and 72, respectively.

In case of FF transitions, as the shape factor $C(w)$ depends on electron
energy, 
it is more appropriate to discuss the averaged shape factor instead of $K_0$.   
Following Ref. \cite{W1}. the averaged shape factor is defined as 

\begin{equation}
\overline{C(w)} =f/f_0
\end{equation}
with
\begin{equation}
f_0 = \int_{1}^{w_0} F(Z,w) pw (w_0-w)^2 dw
, 
\end{equation}
and for FF transitions
\begin{equation}
\overline{C(w)} = \frac{9195\times 10^5}{f_{0}t} \quad ({\rm fm}^2).
\end{equation}

Averaged shape factors for 1$^{-}$ transitions are shown in Fig. 2
(lower figures). Here the matrix elements are multiplied by the electron 
Compton wave length so that they have dimension of 'fm'.   
The transitions have components from the hadronic vector current also, that is, 
the contributions from the electric dipole ($E1$) transitions. 
We see from Fig. 2 that the the total contributions 
have similar strength distributions as those of the spin-dipole part, while 
the contributions from the spin independent part are large.  
The peak position of the transition strength in nuclei near $Z\sim$ 70 is 
located at $\beta$-decay energies of about 6 MeV (4 MeV) for the 1$^{-}$ (GT) 
transition, which is found to be similar to the case of the CQRPA calculation 
in Ref. \cite{Bor03}.

\subsection{Half-lives of the Waiting-point Nuclei} 

Decay rates and relevent matrix elements for the GT and FF transitions 
are evaluated \cite{WB} by including the quenching of the axial vector 
coupling constant: $g_{A}/g_{A}^{free}$=0.7 is taken in the present study
for both the GT and FF transitions except for the 0$^{-}$ case. 
In relation to this, similar order of quenching is found for the spin $g$ 
factor, $g_s^{eff}$ =0.64 $g_s$, in the study of spin-dipole M2 transitions 
in heavy nuclei \cite{Cosel}.
We, therefore, assume the same quenching in the spin-dipole transitions
for 1$^{-}$ and 2$^{-}$ cases as in the GT transitions. 

As for the 0$^{-}$ case, $g_A$ in $M_{0}^{T}$ ($\vec{\sigma}\cdot\vec{p}$
term from $\gamma_5$) is enhanced due to the meson exchange current effects
\cite{W1}. The enhancement factor is taken to be $\epsilon$ =$g_A/g_A^{free}$ 
=2.0 while it is taken to be 1.0 in $M_0^{S}$ ($\vec{\sigma}\cdot\vec{r}$ term)
following the analysis in Ref. \cite{W1}, where $\epsilon$ =2.01$\pm$0.05
and 0.97$\pm$0.06 are obtained for $M_0^{T}$ and $M_0^{S}$, respectively,
by fitting to the experimental $\beta$-decay data for $A$ =205 $\sim$212. 

\begin{table*}[tbh]
\caption{\label{tab:table1}
Calculated half-lives (in units of ms) for the $\beta$-decays of $N$ =126 
isotones obtained by shell model calculations with the use of the quenching 
factor of $g_A/g_A^{free}$ =0.7 for $1^{+}$, $1^{-}$ and $2^{-}$ transitions. 
The enhancement factor of $\epsilon$ =2.0 is used for the matrix element 
$M_0^{T}$ in $0^{-}$ transitions. 
Results for the GT transitions as well as those for both the GT and FF 
transitions are shown for $Z$ = 64$\sim$73.  
Results for the case of the approximation, $\mu_1$ =$\lambda_2$ =1, in the
FF transitions are also shown. 
}
\begin{tabular}{l|cccccccccc}
\hline
Z & 64 & 65 & 66 & 67 & 68 & 69 & 70 & 71 & 72 & 73 \\ 
\hline
 GT (ms) & 5.76 & 7.69 & 11.26 & 17.46 & 29.31 & 54.77 & 102.49 & 223.26 & 
504.79 & 1584.4 \\ 
 GT + FF (ms) & 4.02 & 5.31 & 7.75 & 10.94 & 18.11 & 29.49 & 44.18 & 84.81 & 
129.65 & 278.88 \\ 
 GT + FF ($\lambda_2$=1, $\mu_1$=1) (ms) & 4.01 & 5.31 & 7.74 & 10.94 & 18.08 & 29.48 & 44.07 & 84.73 &
129.24 & 278.35 \\
\hline
\end{tabular}
\end{table*}

\begin{figure*}[tbh]
\hspace{-8mm}
\includegraphics[scale=1.05]{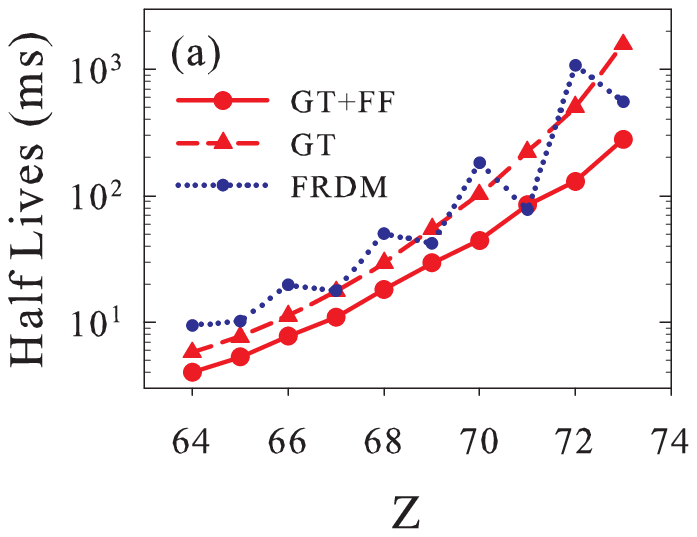}
\hspace*{-12mm}
\includegraphics[scale=1.05]{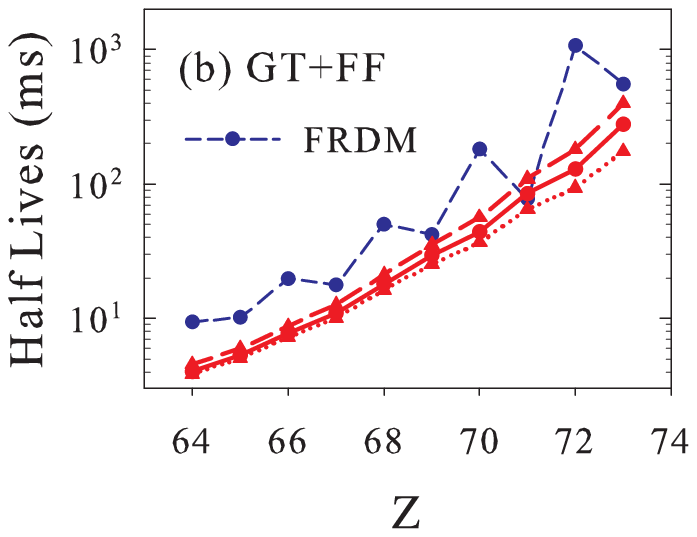}
\caption{(Color online) (a) Calculated half-lives for the $N$=126 isotones.
Results of the present shell model calculations with GT and with GT+FF
transitions are denoted by dashed and solid curves, respectively. 
The quenching factor of $g_A/g_A^{free}$ =0.7 is used for both the GT and
FF transitions except for $0^{-}$ transitions ( see text).  
Half-lives of Ref. \cite{Moll} denoted as FRDM are shown by a dotted curve.
(b) The same as in Fig. 3(a) for the shell model calculations with 
contributions from both the GT and FF transitions. 
The solid curve and the
curve denoted as FRDM are the same as in (a), while the long-dashed and
dotted curves are obtained by using a different quenching factor of 
$g_A/g_A^{free}$ =0.34 with further quenching of $g_V/g_V^{free}$ =0.67.
In case of the dotted curve, the $Q$-values are increased by 1 MeV for
the isotones in the FF transitions.    
\label{fig:fig3}}
\end{figure*}

Calculated half-lives for the $\beta$-decays of the isotones are shown
in Table I and Fig. 3(a).
Calculated $\beta$-decay $Q$-values obtained in the shell model calculations
are used.  
Calculated half-lives obtained with an approximation of using $\mu_1$ =$\lambda_2$
=1 are also shown in Table I. This approximation changes the half-lives by only
within 0.3$\%$. (The effects on the FF transition rates are within 0.5$\%$.)
As the approximation proves to be quite accurate even for the present large $Z$ 
cases, we adopt $\mu_1$ =$\lambda_2$ =1 hereafter.      
The validity of this approximation is also pointed out in Ref. \cite{W1} for
non-unique FF transitions.

\begin{table*}[tbh]
\caption{\label{tab:table2}
Values of log $f_{0}t$ and $\sqrt{\overline{C(w)}}$ for FF transitions in
nuclei with $Z$ =79, 80 and $N$ =126. Calculated values obtained by
using quenching factors of ($g_A/g_A^{fee}$, $g_V/g_V^{free}$) for 
$1^{-}$ transitions and enhancement factor $\epsilon$ for $0^{-}$ 
transitions as well as experimental values\cite{Nudat} are shown. 
Values of Ref. \cite{W1} are also shown for $Z$ =80.
Values of log $f_{0}t$ and $\sqrt{\overline{C(w)}}$ in the parentheses are for
the case in which with proton holes are restricted to 0h$_{11/2}$, 1d$_{5/2}$ 
and 2s$_{1/2}$ orbits.     
}
\renewcommand{\arraystretch}{0.50}
\begin{tabular}{cc|c|c|c|c}
\hline
\multicolumn{2}{c|} {Transitions} & \multicolumn{2}{c|} 
{Values of $g_A$ and $g_V$} &
log $f_{0}t$ &  $\sqrt{\overline{C(w)}}$ (fm)\\
\cline{1-4}
Initial & Final & $\epsilon$ for $0^{-}$ & ($g_A/g_A^{free}$, $g_V/g_V^{free}$)
for $1^{-}$ & & \\
\hline
$^{206}$Hg (0$^{+}$) & $^{206}$Tl (0$^{-}$ g.s.) & 2.0 & & 5.199 (5.087) & 76.3 (86.8)  \\
 & & 1.8 & & 5.432 (5.320) & 58.3 (66.3) \\
 & & Ref. \cite{W1} & & 5.173 & 78.6 \\
 & Expt. \cite{Nudat} & & & 5.41 & 59.8 \\  
\hline
$^{206}$Hg (0$^{+}$) & $^{206}$Tl (1$^{-}$, 0.3049 MeV) &  & (a) (0.34, 0.67) & 5.017 (4.929) & 94.0 (104.1)\\
 & & & (b) (0.51, 0.30) & 5.157 (5.127) & 80.0 (82.9)\\
 & & & (c) (0.47, 0.64) & 4.921 (4.832) & 105.0 (116.4)\\
 & & &     (0.34, 0.40) & 5.267 (5.178) & 70.5 (78.2)\\
 & & &  Ref.\cite{W1} & 5.181 & 77.9 \\
 & Expt.\cite{Nudat} & & & 5.24 & 72.7\\
\hline
$^{205}$Au (3/2$^{+}$) & $^{205}$Hg (1/2$_1^{-}$) & & (a) (0.34, 0.67) & 6.197 (6.116) & 24.2 (26.5)\\
 & & & (b) (0.51, 0.30) & 8.171 (7.834) & 2.49 (3.67)\\
 & & & (c) (0.47, 0.64) & 6.412 (6.326) & 18.9 (20.8)\\
 & & &     (0.34, 0.94) & 5.793 (5.726) & 38.7 (41.6)\\
 &  Expt.\cite{Nudat} & & & 5.79 & 38.6\\
\hline
$^{205}$Au (3/2$^{+}$) & $^{205}$Hg (3/2$_1^{-}$, 0.4675 MeV) & & & & \\
 & 0$^{-}$ & 2.0 & & 5.674 (5.541) & 44.1 (51.5)\\
 & & 1.45 & & 6.832 (6.699) & 11.6 (13.6)\\
 & 0$^{-}$ + 1$^{-}$ & 1.45 & (a) (0.34, 0.67) & 6.502 (6.247) & 17.0 (22.8)\\
 & & 1.45 & (b) (0.51, 0.30) & 6.000 (5.870) & 30.3 (35.2)\\
 & & 1.45 & (c) (0.47, 0.64) & 6.266 (6.073) & 22.3 (27.9)\\
 & & 1.45 & (0.34, 0.94) & 6.425 (6.173) & 18.6 (24.9)\\
 &  Expt.\cite{Nudat} &  &  & 6.43 & 18.5\\
\hline
$^{205}$Au (3/2$^{+}$) & $^{205}$Hg (5/2$_1^{-}$, 0.3792 MeV) & & (a) (0.34, 0.67) & 5.108 (5.006) & 84.7 (95.2)\\
 & & & (b) (0.51, 0.30) & 5.409 (5.308) & 59.9 (67.3)\\
 & & & (c) (0.47, 0.64) & 5.057 (4.956) & 89.8 (100.9)\\
 & & &     (0.10, 0.15) & 6.343 (6.242) & 20.4 (23.0)\\
 &  Expt.\cite{Nudat} & & & 6.37 & 19.8\\
\hline
$^{205}$Au (3/2$^{+}$) & $^{205}$Hg (3/2$_2^{-}$, 1.2806 MeV) & & & & \\
 & 0$^{-}$ + 1$^{-}$ & 1.5 & (a) (0.34, 0.67) & 5.497 (5.457) & 54.1 (56.7)\\
 & & 1.5 & (b) (0.51, 0.30) & 5.556 (5.454) & 50.6 (56.8)\\
 & & 1.5 & (c) (0.47, 0.64) & 5.433 (5.413) & 58.3 (59.6)\\
 & & 1.5 & (0.34, 0.94) & 5.364 (5.374) & 63.0 (62.4)\\
 &  Expt.\cite{Nudat} &  &  & 5.51 & 53.3\\
\hline
$^{205}$Au (3/2$^{+}$) & $^{205}$Hg (5/2$_2^{-}$, 1.2806 MeV) & & (a) (0.34, 0.67) & 5.397 (5.807) & 60.7 (37.9)\\
 & & & (b) (0.51, 0.30) & 5.646 (6.064) & 45.6 (28.2)\\
 & & & (c) (0.47, 0.64) & 5.333 (5.746) & 65.4 (40.6)\\
 & & &     (0.34, 0.94) & 5.176 (5.584) & 78.3 (48.9)\\
 &  Expt.\cite{Nudat} & & & 5.51 & 53.3\\
\hline
$^{205}$Au (3/2$^{+}$) & $^{205}$Hg (1/2$_2^{-}$, 1.4472 MeV) & & (a) (0.34, 0.67) & 6.632 (6.163) & 14.7 (25.1)\\
 & & & (b) (0.51, 0.30) & 7.191 (7.956) & 7.7 (3.2)\\
 & & & (c) (0.47, 0.64) & 6.708 (6.386) & 13.4 (19.4)\\
 & & &     (0.34, 0.94) & 6.300 (5.767) & 21.5 (39.7)\\
 &  Expt.\cite{Nudat} & & & 6.29 & 21.7\\
\hline

\end{tabular}
\end{table*}      

Calculated half-lives obtained by the GT contributions only are found to be  
close to those in Refs. \cite{Lang,GLM,Kurt} within $\sim$ 10$\%$ 
except for $Z$ =71 and 73. In case of $Z$ =71 and 73, the half-lives obtained
here are shorter by about 1.3$\sim$1.5 than those in Refs. \cite{Lang,GLM}. 
The shell model interaction used here\cite{Stee} is not the same as that
in Refs. \cite{Lang,GLM}.    
The half-lives become shorter for isotones with smaller $Z$ as both the strength
and the energy difference between initial and final states get larger. 
The FF transitions are found to be important to reduce the half-lives
by 1.4$\sim$5.7 times of those given by the GT contributions only. 
Their contributions become more important compared to the GT ones for larger 
$Z$ cases.
The dominant contributions come from the 1$^{-}$ transitions. 
The contributions from the $0^{-}$ transitions are about 10$\%$ of the 
total FF contributions; 7$\sim$8 $\%$ for $Z$ =64$\sim$68, 9$\sim$10 $\%$
for $Z$ =69$\sim$71 and 11$\%$ (13$\%$) for $Z$ =72 (73). 
The reduction of the half-lives due to the inclusion of the FF transitions
is not as large as that obtained in Ref. \cite{Bor03}. 
The half-lives obtained here are, however, short compared with the standard
data of Ref. \cite{Moll} except for $Z$ =71 usually employed in 
nucleosynthesis network 
calculations as shown in Fig. 3(a). 
The present half-lives of the shell model 
calculations are shorter than those of the standard values 
by 2.3$\sim$8.3 for even $Z$ and by 1.4$\sim$2.0 for odd $Z$ (except for
$Z$ =71), respectively.
They increase monotonically as $Z$ increases showing no
odd-even staggering found in FRDM's. 
This is due to the absence of the odd-even staggering in the $\beta$-decay 
$Q$-values in the shell model calculations in contrast to the case of FRDM
calculations.

The present half-lives are longer than those of Ref. \cite{Bor03} 
by about 1.1$\sim$1.3 (1.5) for $Z$ =64$\sim$67 (68) and by twice for 
$Z$ =69 and 70, respectively.
They are, on the other hand, short compared to the half-lives of Ref. 
\cite{Kurt}, where similar CQRPA calculations have been done as
in Ref. \cite{Bor03} but without energy dependent smearing of the GT and FF
transition strengths. The present half-lives are shorter than those of
Ref. \cite{Kurt} by about a factor of 1.5 for $Z$ =69, 70 and about by twice 
for $Z$ =
72, 73, while they are close to each other for $Z$ =64$\sim$68 and 71.
As the shell-model calculations include the spreading of the
GT and FF transition strengths, it is reasonable that shorter half-lives
are obtained in the present calculations.

\begin{table*}[tbh]
\caption{\label{tab:table3}
Half-life for the $\beta$-decay of $^{204}$Pt.  Calculated values obtained 
by using quenching factors of ($g_A/g_A^{fee}$, $g_V/g_V^{free}$) for 
$1^{-}$ transitions and enhancement factor $\epsilon$ for $0^{-}$ transitions
as well as experimental value \cite{Nudat} are shown. 
Values of the half-life in the parentheses are for
the case in which proton holes are restricted to 0h$_{11/2}$, 1d$_{5/2}$ 
and 2s$_{1/2}$ orbits.    . 
}
\begin{tabular}{cc|c|c|c}
\hline
\multicolumn{2}{c|} {Transitions} & \multicolumn{2}{c|} 
{Values of $g_A$ and $g_V$} &
Half-life (s)\\
\cline{1-4}
Initial & Final & $\epsilon$ for $0^{-}$ & ($g_A/g_A^{free}$, $g_V/g_V^{free}$)
for $1^{-}$ &  \\
\hline
$^{204}$Pt (0$^{+}$) & $^{204}$Au (0$^{-}$) & 2.0 & & 65.1 (62.0) \\
\hline
$^{204}$Pt (0$^{+}$) & $^{204}$Au ($0^{-}$ + 1$^{-}$) & 2.0 & (a) (0.34, 0.67) & 22.8 (18.6)\\
 & & 2.0 & (b) (0.51, 0.30) & 31.8 (26.9)\\
 & & 2.0 & (c) (0.47, 0.64) & 21.1 (17.1)\\
 & & 2.0 &     (0.7, 1.0) & 10.9 (8.6)\\
 & & 2.0 & (0.34, 1.0) & 14.4 (11.3) \\
 & & 2.0 & (0.51, 1.0) & 12.7 (10.0)\\
 & Expt.\cite{Nudat} & & & 10.3$\pm$1.4 \\
\hline
\end{tabular}
\end{table*}

We here consider other possible quenching factors for 1$^{-}$
transitions. A large quenching of $g_A$ and $g_V$ for 1$^{-}$ spin-dipole
transitions was suggested by studies of FF $\beta$-decays and related processes 
in the lead region \cite{W2,W1,Ryd}. 
In Ref. \cite{W2}, effective quenching factors of $g_A/g_A^{free}\sim$0.47 and 
$g_V/g_V^{free}\sim$0.64 due to core polarization effects are obtained for 
1$^{-}$ transitions.
In Ref. \cite{Ryd}, two sets of quenching factors, $g_A/g_A^{free}$ =0.34,
$g_V/g_V^{free}$=0.67 and $g_A/g_A^{free}$ =0.51, $g_V/g_V^{free}$ =0.30,
are obtained from the analysis of the FF transition, $^{205}$Tl (1/2$^{+}$,
g.s.) $\rightarrow$ $^{205}$Pb (1/2$^{-}$). 

We study FF $\beta$-decays in nuclei with $Z$ =78$\sim$80 and $N$ =126,
where experimental data are available for $0^{-}$ and $1^{-}$ transitions
\cite{Nudat}.

Calculated values of log $f_{0}t$ and $\sqrt{\overline{C(w)}}$ (fm) for the FF 
$\beta$-decays, $^{206}$Hg $\rightarrow$ $^{206}$Tl ($0^{-}$, $1^{-}$)
and $^{205}$Au $\rightarrow$ $^{205}$Hg (1/2$^{-}$, 3/2$^{-}$, 5/2$^{-}$), 
are given in Table II for the following sets of the quenching factors;
($g_A/g_A^{free}$, $g_V/g_V^{free}$) = (a) (0.34, 0.67) \cite{Ryd}, (b) 
(0.51, 0.30) \cite{Ryd}
and (c) (0.47, 0.64) \cite{W2} for 1$^{-}$ transitions. Experimental 
values \cite{Nudat} as well as the calculated values given in Ref. \cite{W1} 
for $Z$ =80 and cases for some other sets of quenching such as (0.34, 0.40)
and (0.34, 0.94) are also given. 
In case of $0^{-}$ transitions, the enhancement factor of $\epsilon$ =
$g_A/g_A^{free}$ =1.45$\sim$2.0 is adopted for the matrix element, $M_0^{T}$.
Values in the parentheses in the Table  are for the case in which proton
holes are restricted to 0h$_{11/2}$, 1d$_{3/2}$ and 2s$_{1/2}$ orbits.

In case of $Z$ =80, for the $0^{-}$ transition, calculated log $f_{0}t$ and
$\sqrt{\overline{C(w)}}$ values for $\epsilon$ =2.0 are close to those of Ref. \cite{W1}
and become closer to the experimental values for $\epsilon$ =1.8. 
For the $1^{-}$ transition, the calculated log $f_{0}t$ values for the sets (a), (b) and 
(c) are smaller than those of Ref. \cite{W1} by 0.17, 0.02 and 0.26, respectively. 
They get closer to
the observed values if a set with more quenching for $g_V$, (0.34, 0.40), is
used. 

In case of Z=79, transitions to the 1/2$_{1}^{-}$ and 5/2$_{1}^{-}$ states
in $^{205}$Hg are dominantly induced by 1$^{-}$ transitions, while the
transition to the 3/2$_{1}^{-}$ state is a mixture of $0^{-}$ and $1^{-}$
transitions. For the $1^{-}$ transition to the 1/2$_{1}^{-}$ (g.s.) state, the
three sets of the quenching give larger (smaller) log $f_{0}t$ 
($\sqrt{\overline{C(w)}}$) values than the experimental ones. 
A set with smaller quenching for $g_V$, (0.34, 0.94), gives the values
close to the experiment. For the transition to the 3/2$_{1}^{-}$ state 
($E_x$ =0.4675 MeV), the $0^{-}$ transition proceeds too fast for 
$\epsilon$ =2.0. 
When an enhancement factor of $\epsilon$ =1.45 is used, the three sets of
the quenching for the $1^{-}$ transitions can give log $f_{0}t$ and
$\sqrt{\over{C(w)}}$ values close to the experimental ones.
The set used for the 1/2$_1^{-}$ case, (0.34, 0.94), works also in this 
transition. 
As for the 5/2$_{1}^{-}$ state, any of the three sets of the quenching for the 
$1^{-}$ transition give transitions too fast compared to the experiment. 
In this case, an exceptionally large quenching, (0.10, 0.15), is necessary 
to explain the experiment. 

Although the spins of the excited states of $^{205}$Hg at $E_x$ =1.2806 MeV
and $E_x$ =1.4472 MeV are not determined yet, the present analysis suggests
that the state at $E_x$ =1.2806 MeV is 3/2$^{-}$ or 5/2$^{-}$ and the state 
at $E_x$ =1.4472 MeV is 1/2$^{-}$ (see Table II). 
Other possibilities, that is, 1/2$^{-}$ (3/2$^{-}$ and 5/2$^{-}$)
for the state at $E_x$ =1.2806 MeV (1.4472 MeV) can be excluded.

In case of $Z$ =78, the half-life for the transition, $^{204}$Pt (0$^{+}$)
$\rightarrow$ $^{204}$Au has been obtained by measuring two $\gamma$ decays in 
$^{204}$Au \cite{Nudat}. The averaged experimental half-life is 10.3 s.
Here, the half-life is calculated by taking into account the FF $0^{-}$ and
$1^{-}$ transitions. The enhancement factor of $\epsilon$ =2.0 is used for
the $0^{-}$ transition. The half-lives obtained with the three sets of
the quenching for the $1^{-}$ transitions are given in Table III. 
The calculated half-lives are longer than the observed one by about 2$\sim$3.
The values become close to the observed one for the cases without quenching 
for $g_V$ (see Table III). 
The present analysis suggests smaller quenching for $g_V$ for $Z$ =78.

The present analysis for $Z$ =78$\sim$80 suggests a large quenching for $g_A$.
A large quenching for $g_V$ is not necessarily needed for $Z$ =78 and 
for most cases of $Z$ =79.
We study dependence of the calculated half-lives on the quenching of
$g_A$ and $g_V$. Half-lives obtained with the set (a) for the $1^{-}$
transitions are shown in Fig. 3(b). The enhancement factor $\epsilon$ for
the $0^{-}$ transitions is not changed; $\epsilon$ =2.0.
The half-lives get longer but still remain shorter than those of Ref. \cite
{Moll} except for $Z$ =71. They become close to the values of Ref. \cite{Kurt}. 

Although the experimental $Q$-values are used for the analysis of the
$\beta$-decays for $Z$ =78$\sim$80, the present shell-model calculations
give smaller $Q$-values compared to the experimental ones.
The differences are 1.3, 1.6 and 0.9 MeV for $Z$ =80, 79 and 78, respectively.
As there are no experimental information on the $Q$-values for $Z$ =64$\sim$73,
we used the calculated $Q$-values. 
They are small compared to the mass differences between the parent 
and daughter nuclei in Ref. \cite{Moll} by about 2 MeV (3 MeV) for even (odd)
$Z$ nuclei.
Now, we study the dependence of the half-lives on the $Q$-values of the
transitions. 
Calculated half-lives are shown in Fig. 3(b) for the case, in which
the $Q$-values obtained by the shell model calculations are increased by 
1 MeV for the isotones with $Z$ =64$\sim$73.
The effects of the change of the $Q$-values, which results in the enhancement
of the phase space volumes, are found to be large. 
The half-lives get shorter than the case in Fig. 3(a) with the quenching, (0.7, 1.0),
but they do not become as short as those of Ref. \cite{Bor03}. They are still
longer than those of Ref. \cite{Bor03} by about 30$\%$ for $Z$ =66$\sim$68 and 
50$\%$ (60$\%$) for $Z$ =69 (70). 
It is quite important to obtain experimental information on  
the masses of waiting point nuclei and their daughters.

\section{R-process Nucleosynthesis}

\begin{table*}[tbh]
\caption{\label{tab:table4}
The mass shift at the third peak of the element abundance in the r-process
defined by eq. (8) as well as the peak position for the case of the 
modified half-lives 
are shown for several astrophysical conditions.
The ratio (R) of the height of the second peak over that of the third peak
is also given. For each astrophysical condition, the first row corresponds
to the case of the quenching of ($g_A/g_A^{free}$, $g_V/g_V^{free}$)
=(0.70, 0.0), while the second and the third rows correspond to the case of
(a) (0.34, 0.67) for 1$^{-}$ tarnsitions. The $Q$-value is increased by 1 MeV 
for the third row (denoted as (a)$^{\ast}$).    
}
\begin{tabular}{c|c|c|c|c|rcc}
\hline
$L_{\nu,51}$ & $S/k$ & $\dot{M}$ ($M_\odot$ s$^{-1}$) & $f_t$ & 
$\tau$ (ms) & ratio R & $<A>_{{\rm Mod}}$ & $\Delta A$\\
\hline
0.5 & 133.4 & $2.34 \times 10^{-6}$ & 0.06 & 5.60 & 3.45 & 196.972 & 1.101 \\ 
 & & & & & (a)$^{~}$ 3.21 & 196.684 & 0.813\\
 & & & & & (a)$^{\ast}$ 3.84 & 197.218 & 1.347 \\
\hline
1.0 & 118.9 & $7.43 \times 10^{-6}$ & 0.08 & 4.04 & 4.23 & 197.118 & 1.058 \\
 & & & & & (a)$^{~}$ 3.89 & 196.837 & 0.777 \\
 & & & & & (a)$^{\ast}$ 4.75 & 197.359 & 1.299 \\
\hline
2.0 & 105.9 & $2.36 \times 10^{-5}$ & 0.10 & 2.90 & 4.77 & 197.234 & 1.008 \\
 & & & & & (a)$^{~}$ 4.35 & 196.965 & 0.739 \\
 & & & & & (a)$^{\ast}$ 5.03 & 197.466 & 1.240 \\
\hline
5.0 & 90.91 & $1.08 \times 10^{-4}$ & 0.11 & 1.78 & 3.80 & 197.313 & 0.949 \\ 
 & & & & & (a)$^{~}$ 3.52 & 197.066 & 0.702 \\
 & & & & & (a)$^{\ast}$ 3.94 & 197.523 & 1.159 \\
\hline
\end{tabular}
\end{table*}

\begin{figure*}[tbh]
\includegraphics[scale=0.45]{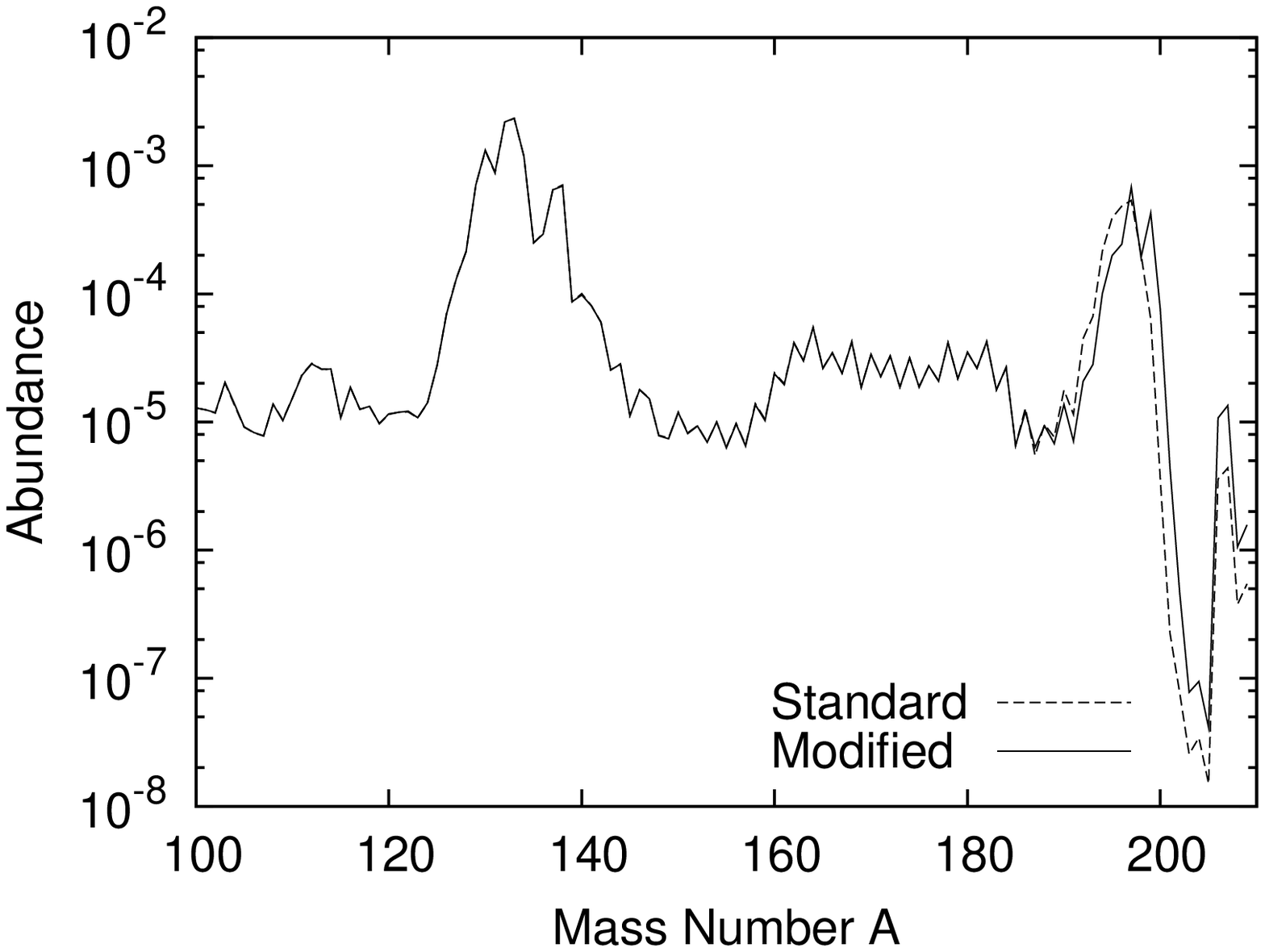}
\hspace*{-3mm}
\includegraphics[scale=0.45]{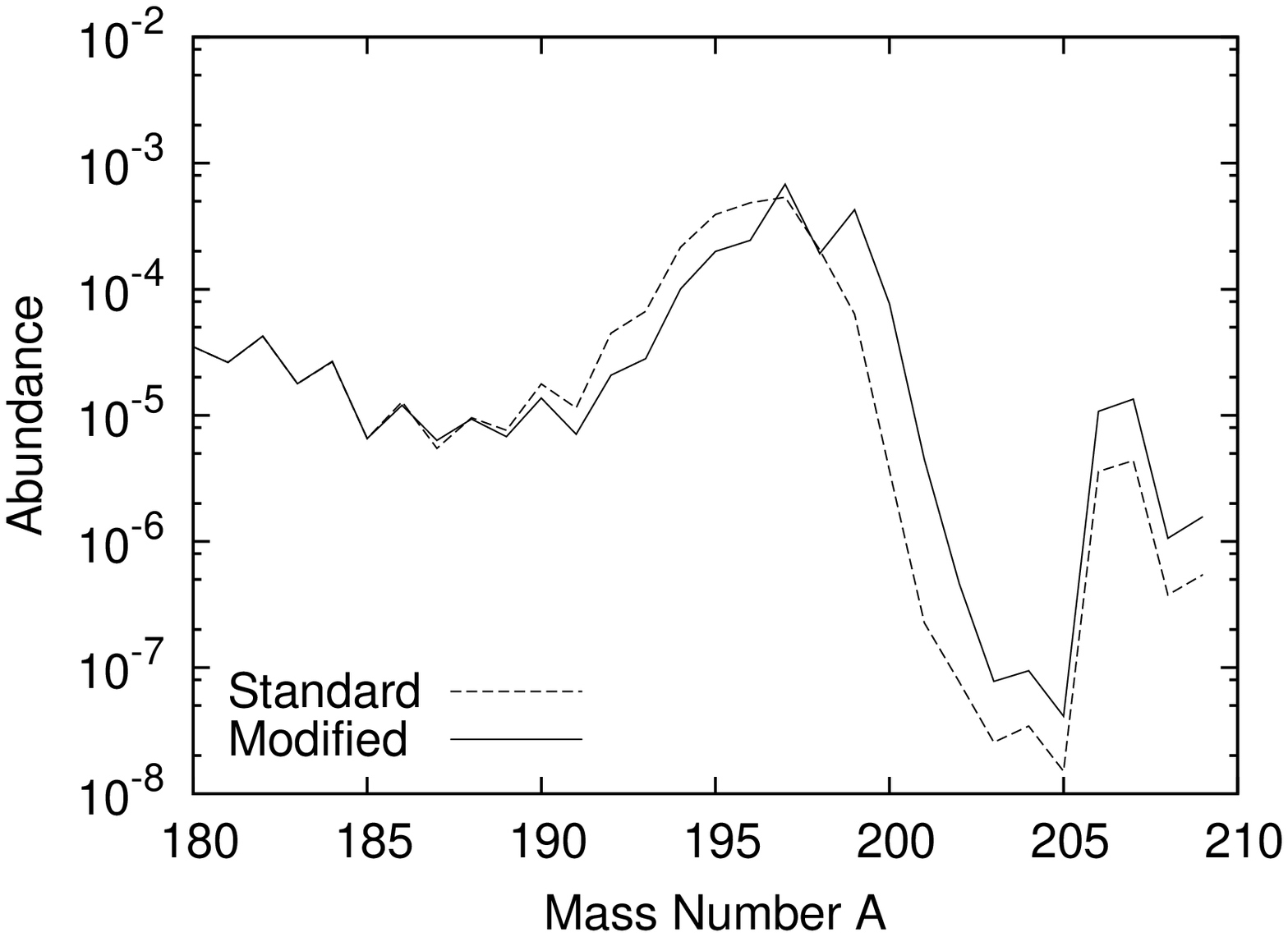}
\caption{
The abundances of elements in the r-process nucleosynthesis 
obtained by using the present $\beta$-decay half-lives for the $N$=126
isotones (denoted as 'modified') and standard half-lives of 
ref. \cite{Moll} (denoted as 'standard').
The quenching factor of $g_A/g_A^{free}$ = 0.7 ($g_V/g_V^{free}$ =1.0) 
is used both for the GT and FF ($1^{-}$ and $2^{-}$) transitions.
The right figure displays the same result in the restricted range of $A\ge$180.  
\label{fig:fig4}}
\end{figure*} 

\begin{figure*}[tbh]
\includegraphics[scale=0.45]{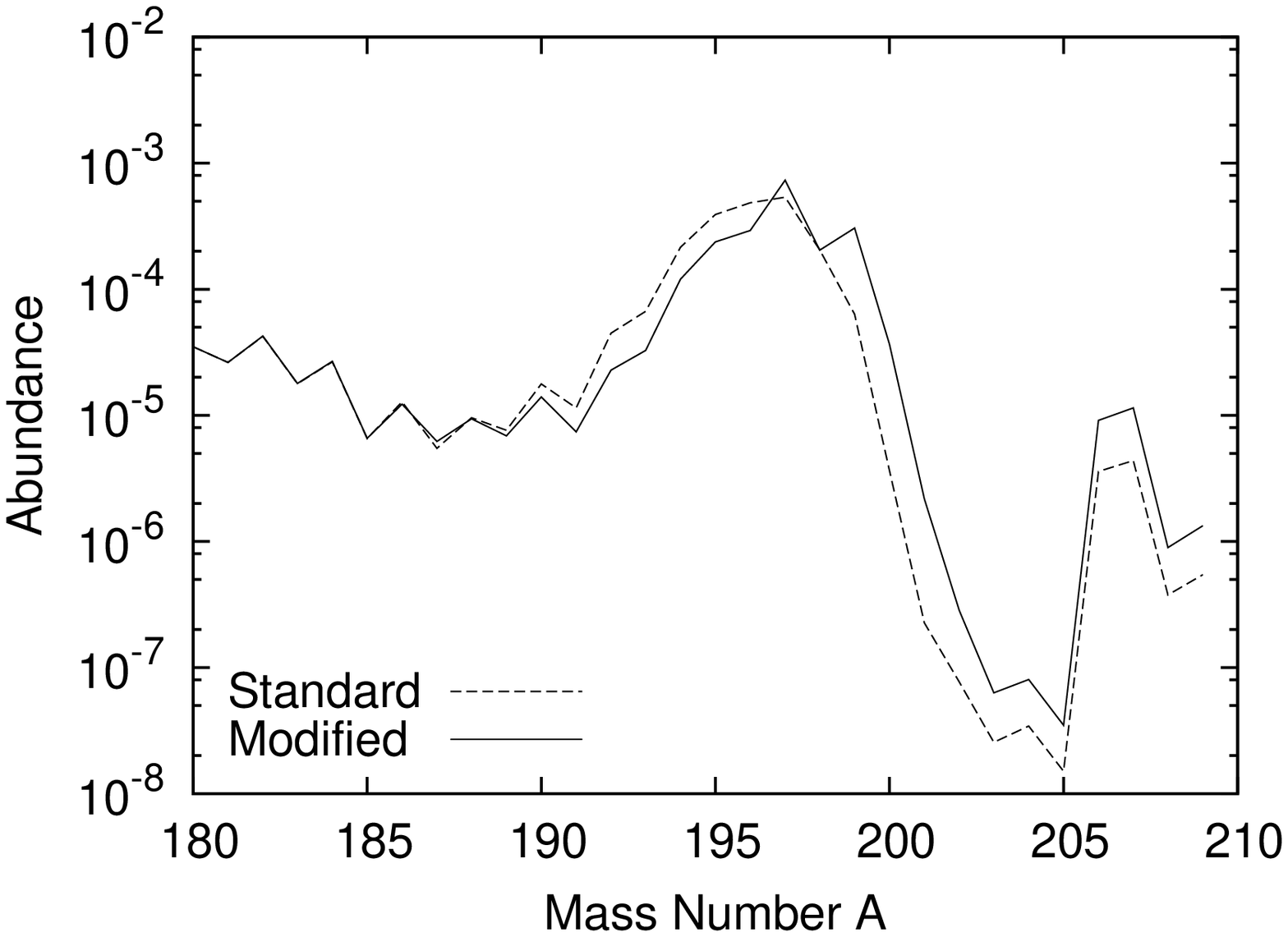}
\hspace*{-3mm}
\includegraphics[scale=0.45]{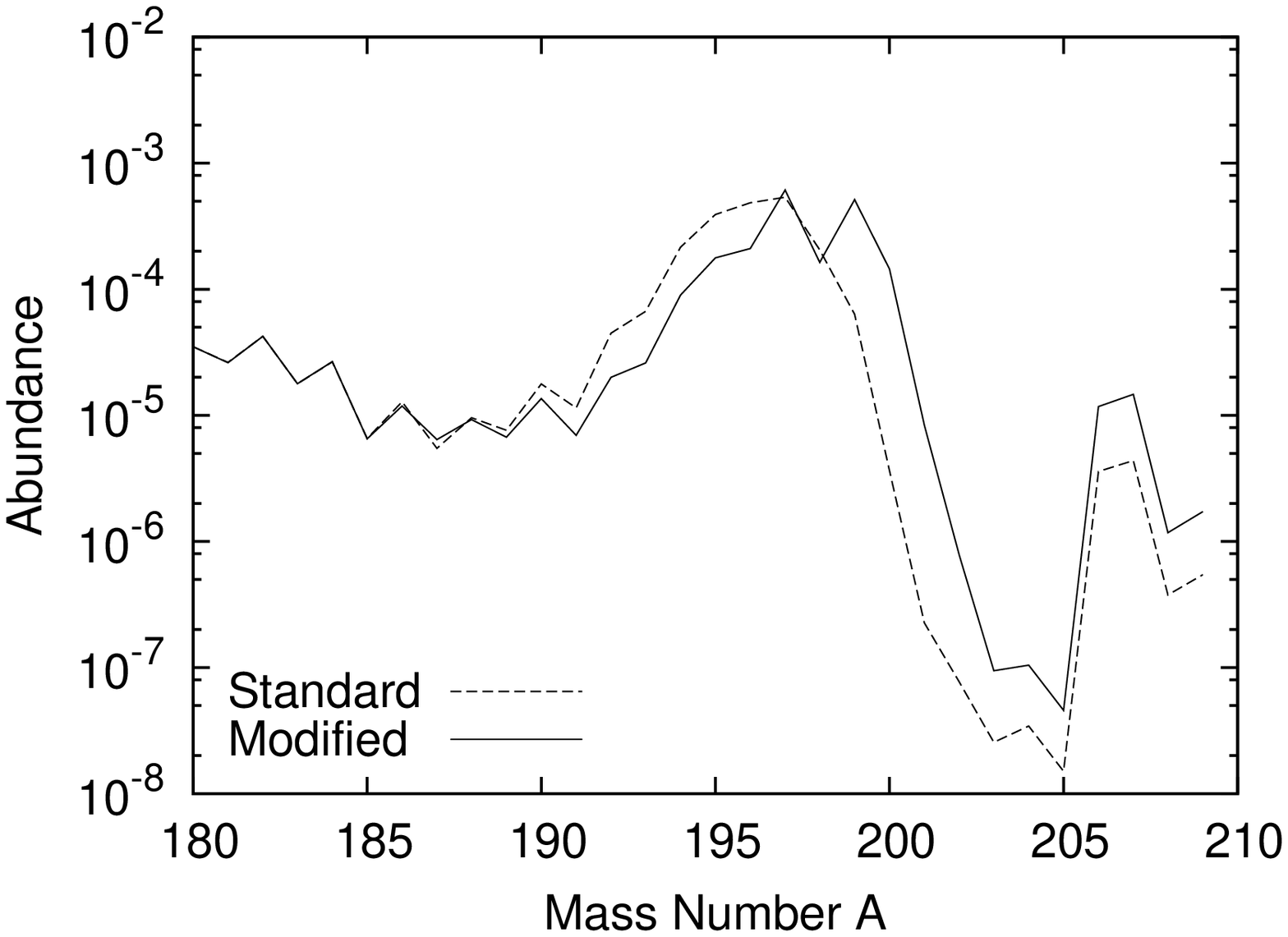}
\caption{
The same as in Fig. 4 at $A\ge$180 for the quenching factors of 
($g_A/g_A^{free}$, $g_V/g_V^{free}$)  =(0.34, 0.67) for the $1^{-}$ 
transitions.  
The figure on the right side includes the effect of the increase of the
$Q$-values.  
\label{fig:fig5}}
\end{figure*}

We discuss possible effects of the short half-lives of the waiting 
point nuclei at $N$=126 obtained in the preceding section 
on the r-process nucleosynthesis.
The dependence of the abundances of the elements around mass number 
A$\sim$195 on the half-lives of the nuclei is investigated for 
various astrophysical conditions.
We use an analytic model for neutrino-driven winds \cite{Tak} for
the time evolution of thermal profiles.
The wind solutions which pass the sonic points are adopted.
We take into account the relation of the neutrino luminosity $L_\nu$ and 
the mean neutrino energy to the entropy per baryon $S/k$ and the mass ejection 
rate $\dot{M}$ of the winds in accordance with \cite{Qian}.
The neutrino energy spectra are assumed to obey Fermi distributions with
zero chemical potentials.
The temperatures of $\nu_e$, $\bar{\nu}_e$, and $\nu_x = \nu_{\mu,\tau}$ and
$\bar{\nu}_{\mu,\tau}$ are set to be
$(T_{\nu_e}, T_{\bar{\nu}_e}, T_{\nu_x})$ = (3.2 MeV, 5 MeV, 6 MeV)
\cite{TY04}.
The luminosity of each flavor of neutrinos is equipartitioned and is
taken to be $L_\nu = (0.5 \sim 5.0) \times 10^{51}$ erg s$^{-1}$.
The neutrino luminosity, entropy per baryon, and mass ejection rate
of each wind model are listed in Table IV.

We calculate r-process nucleosynthesis using a reaction network 
consisting of 3517 species of nuclei.
The initial electron to baryon ratio is taken to be $Y_e$ =0.40. 
Neutrino processes on nucleon and $^{4}$He are included in the 
network calculations.
Charged current reactions, n($\nu_e, e^{-}$)p, p($\bar{\nu}_e, e^{+}$)n,
$^{4}$He ($\nu_e, e^{-}$p) $^{3}$He, and 
$^{4}$He ($\bar{\nu}_e, e^{+}$n) $^{3}$H give dominant contributions.
Besides them, charged current reactions, $^{4}$He ($\nu_e, e^{-}$pp) $^{2}$H, 
$^{4}$He ($\bar{\nu}_e, e^{+}$nn) $^{2}$H,
and neutral current reactions, $^{4}$He ($\nu, \nu'$n) $^{3}$He, 
$^{4}$He ($\nu, \nu'$p) $^{3}$H, $^{4}$He ($\nu, \nu'$d) $^{2}$H, 
$^{4}$He ($\nu, \nu'$nnp) $^{1}$H and those induced by $\bar{\nu}$'s are
included. 
Cross sections of ref. \cite{SCY} obtained by the WBP Hamiltonian 
\cite{WBH} are used.
The cross sections of n and p are adopted from \cite{Hor}.
The neutrino processes suppress the production of the elements
around the third peak in the r-process as neutrons are changed into
protons, which results in the increase of $Y_e$, and the production of
seed elements is enhanced through the neutrino-induced breakup of $^{4}$He
\cite{Mey}. 
We artificially shorten the time scale of the explosion
by a constant factor $f_t$ in order to proceed the r-process
to the synthesis of the 3rd peak elements 
because it was found that the shorter expansion time scale tends to
decrease the roles of neutrino interactions and keep low $Y_e$ in
favor of the r-process \cite{MMH}.
We set the final temperature of the winds to be $T_f = 8 \times 10^{8}$ K.
The duration time $\tau$ where the temperature decreases from
$5 \times 10^{9}$ K to $2 \times 10^9$ K is 
$\tau = (1.8 \sim 5.6)$ ms.
The multiplying factor $f_t$ and the duration time $\tau$ for r-process
calculation are also shown in Table IV.

The abundance of the elements obtained by using the present short 
half-lives in the r-process element network are compared with 
those obtained with the use of the standard data of ref. \cite{Moll}. 
We show in Fig. 4 a case 
in which the ratio of the height of the third peak
over that of the second peak in the r-process is close to the solar
abundance ratio, 3$\sim$4. 
Parameters chosen are $L_{\nu,51}$ =0.5 
where $L_{\nu,51}$ =$L_{\nu}$/(10$^{51}$ erg s$^{-1}$),  
$S/k$ = 133 and $\tau$ = 5.6 ms.
Here, the half-lives obtained for the case with the quenching of 
$g_A/g_A^{free}$ =0.7 ($g_V/g_V^{free}$ =1) for $1^{-}$ transitions
are used. Other inputs except for the half-lives are not changed.
We show in Fig. 5 the calculated results for the case of a different set of the
quenching factors for $1^{-}$ transitions, ($g_A/g_A^{free}$, $g_V/g_V^{free}$) 
=(0.34, 0.67). A case with $Q$-values increased by 1 MeV is also shown.  
We find a slight shift of the third peak of the element abundances toward 
higher mass region. 
This shows more rapid build-up of the r-process elements caused by the 
shorter half-lives of the waiting point nuclei. The isotones with higher
$Z$ accumulate more abundantly at the freezing time of the neutron-capture
flow, 
which by successive $\beta$-decays naturally results in larger abundances 
of the elements at higher mass region.

When the decrease of the temperature becomes slower
or the entropy of the system is larger, the magnitude of the shift of 
the peak gets larger.    
The shift of the mass number at the third peak region is defined by

\begin{eqnarray}
\Delta A &=& <A>_{{\rm Mod}} -<A>_{{\rm STD}} \nonumber\\
<A> &=& \frac{\sum_{A} A\cdot Y(A)}{\sum_{A} Y(A)}
\end{eqnarray}

\noindent where $Y(A)$ is the abundance of the element with mass 
number $A$
and the summation is taken over 189$\leq A \leq$203.
'Mod' and 'STD' refer to the cases of modified half-lives and
standard ones, respectively. 
The mass shifts at the third peak of the r-process are shown in
Table IV for several different values of $L_{\nu,51}$, $S/k$ and $\tau$.  
Here, the condition is being kept so that the ratio of the height of the 
third peak over that of the second peak is close to the solar abundances 
ratio.
We see from Table IV that the mass sfift $\Delta A$ is about unity in all 
cases.
Dependence on the astrophysical condition is found to be small.
Dependence on the difference of the quenching factors is also as small
as 25$\sim$30 $\%$. 
The decreace of the mass sfift due to larger quenchings is recovered 
or even reversed to an increase by the change of the $Q$-values.    
We also find that the mass shift at the second peak, where the average
is taken over 124$\leq A \leq$136, is quite small; 
$\Delta A = -0.006 \sim 0.002$.
We thus emphasize that the shift of the third peak of element abundance
toward higher mass region by about unity is a common phenomenon in 
the present astrophysical conditions. 
When half-lives of the $N$ =82 isotones are changed to the values
obtained by shell model calculations \cite{LP}, effects on the shift of 
the third peak is found to be quite small. The mass shift $\Delta A$
decreases only by 6$\%$ while the mass shift at the second peak is 
$\Delta A$ =0.26$\sim$0.28. 
The magnitude of the shift at the third peak is rather modest, but 
it is a robust effect and is of the same order as that 
caused by variations of mass formulae. 
As we see from Table IV, the position of the third peak of the solar 
abundance, $A\sim$190, is not necessarily reproduced by the present 
calculations.  
However, we should note that the third peak region can be affected by 
various effects besides the half-lives of the $\beta$-decays
as we pointed out.
Neutron-star merger and gamma-ray bursts are considered to be possible 
alternative r-process sites \cite{Thi,Fuji06}.  
The assignment of the site of r-process including those besides the 
supernova explosions is an important and interesting subject to be
explored further in future.

\section{Summary}

In summary, we have evaluated half-lives of the $\beta$-decays of $N$ =126 isotones
taking into account both the GT and FF transitions. The half-lives have been
found to be reduced by including the FF transitions. The effects of the short 
half-lives obtained here on the nucleosynthsis of the r-process are investigated.
The third peak of the abundance of the elements in the r-process has been 
found to be shifted toward higher mass region.    
Although the magnitude of the shift is rather modest, it is found to
be a robust effect independent of the present astrophysical conditions for 
the r-process as well as the quenching factors of $g_A$ and $g_V$ adopted
in the shell model calculations.
The results obtained here might be improved quantitatively by overcoming 
the limitations in the present structure calculations such as configuration 
space. 
Improvements of various nuclear properties for the input for the 
nucleosynthesis network of the r-process in addition to the $\beta$-decay 
half-lives, 
especially the masses of nuclei at and around the waiting points, 
are important issues to be made in future studies.\\

The authors would like to thank Y. Utsuno for useful 
discussions. 
This work has been supported in part by Grants-in-Aid for Scientific
Research (C) 20540284, 22540290, 23540287, (A) 20244035, and on Innovative 
Areas (20105004) 
of the Ministry of Education, Culture, Sports,
Science and Technology of Japan, and also by JPSJ Core-to-Core Program,
International Research Network for Exotic Femto Systems (EFES).


\end{document}